# The Boron Buckyball has an Unexpected $T_h$ Symmetry

G. Gopakumar, Minh Tho Nguyen, and Arnout Ceulemans*

*Department of Chemistry and Institute for Nanoscale Physics and Chemistry, University of Leuven, Celestijnenlaan 200F, B-3001 Leuven, Belgium*

**(Abstract):** The boron buckyball avoids the high symmetry icosahedral cage structure. The previously reported $I_h$ symmetric structure is not an energy minimum in the potential energy surface and exhibits a spontaneous symmetry breaking to yield a puckered cage with a rare $T_h$ symmetry. The HOMO-LUMO gap is twice as large as the reported value and amounts to 1.94 eV at B3LYP/6-31G(d) level. The valence orbital structure of boron buckyball is identical to the one in the carbon analogue.

In a recent theoretical study, Szwacki, Sadrzadeh and Yakobson predicted the existence of an unusually stable new boron cage with 80 boron atoms.[1] This cluster was named the *boron buckyball* in view of its similarity with the celebrated $C_{60}$ buckminsterfullerene. As in the carbon structure, 60 boron atoms are situated at the vertices of a truncated icosahedron consisting of pentagons and hexagons, whereas an additional set of 20 borons are located in the centre of each hexagon. This orbit of 20 capping atoms forms a dodecahedron which is superimposed on the truncated icosahedron. A plane-wave density functional theory calculation indicated that this structure has $I_h$ symmetry, and attains a cohesive energy of 5.76 eV/atom, which is a record for boron allotropes. The structure can be built from overlapping boron double-rings that are forming interlaced strips wrapped around the centre of a sphere. In the present contribution we examine the *boron buckyball* more closely, using quantum chemical calculations, and thereby report some unexpected features.

A straightforward geometry optimization using the popular hybrid B3LYP functional of density functional theory, in conjunction with the polarized 6-31G(d) basis set, under $I_h$ symmetry constraint yields a closed-shell singlet cage, with B-B bond lengths of 1.74 and 1.67 Å. The boron caps are nearly coplanar, the dihedral angle with respect to the six-member ring amounts to 3.7°. The calculated binding energy per boron atom amounts to −5.17 eV [−5.82 eV at PBE/SV(P) level]. These results are very similar to the outcome of the calculations reported in ref. [1], with however the following exceptions: i) the HOMO-LUMO energy gap is about twice as large as the reported value of 1.006 eV and corresponds now to 1.94 eV [1.01 eV at PBE/SV(P) level]; ii) moreover the HOMO is a fivefold degenerate orbital with $h_u$ symmetry, in contrast with the claim of threefold degeneracy, and iii) last but not least, when a vibrational frequency analysis was performed, we found that this cage has seven imaginary frequencies, with $t_{2u} + g_g$ representations. The absolute values of the imaginary frequencies are rather small, but significant, ranging from 67$i$ to 96$i$ cm$^{-1}$ at B3LYP/6-31G(d), 100$i$ to 129$i$ cm$^{-1}$ at B3LYP/SV(P) and 79$i$ to 102$i$ cm$^{-1}$ at PBE/SV(P) level. This feature is already found with Hartree-Fock calculations and persists when using different functionals. A conclusion must be imposed that the *predicted structure in ref. [1] is not an energy minimum on the potential energy surface.*

For the present analysis, the preliminary geometry optimizations were carried out at Hartree-Fock level in conjunction with the minimal STO-3G basis set using the Gaussian 03 Revision D02 program package.[2] The stationary points were then reoptimized using both HF and DFT method (in particular the popular B3LYP functional) in conjunction with the larger polarized 6-31G(d) basis set. The located stationary points were characterized by the harmonic vibrational frequency analysis at the same level. The seven imaginary modes of the $I_h$ symmetric structure were reconfirmed at the DFT level using different functionals, including BP86, PBE and B3LYP in conjunction with a split-valence plus polarization SV(P) basis set, which is having the same quality as 6-31G(d) but implemented in the TURBOMOLE V5-8-0 program package.[3] The molecular orbitals were visualized and plotted with the help of gOpenmol program. [4, 5]

When the geometry optimization was continued with no symmetry constraints whatsoever, energy is lowered and a stable energy minimum is found, which exhibits an unexpected $T_h$ symmetry. The resulting structure is visualized as **A** in Fig. 1. This structure was then reoptimized at the B3LYP/6-31G(d) level and the located minimum is characterized by harmonic vibrational frequencies evaluated at the same level (See Table 1 for total and relative energies).

In geometry **A**, the hexagonal $B_6$ rings remain almost planar but 8 of them shrink considerably while the remaining 12 become slightly larger. This is accompanied by a concerted motion of the capping atoms: the 8 boron atoms in the centre of contracting rings move inward towards the centre of the cage, while the 12 remaining ones show a very slight shift in the opposite direction. This corresponds to a $T_h$ distortion of the original cap-dodecahedron. There is also an alternate geometry **B**, where the direction of the distortion coordinate is reversed: while 12 B-atoms are now endohedral, the remaining 8 are in the plane of the hexagons. Table 1 shows the results of calculated energies on both isomers, and Fig. 1 displays the structural information. Due to the size of the basis sets employed here, it seems reasonable to consider that both structures **A** and **B** have comparable energy content. Unfortunately, due to the computational difficulties we were unable to employ correlated methods for the sake of chemical accuracy. The symmetry lowering is confirmed at all levels of theory considered, with a stabilization energy of about -1.0 kcal/mol. In the structure **A**, the endohedral boron atoms are well inside the cage, making a dihedral angle of $-9.04°$ with the six-membered rings; the corresponding centroid-endohedral boron distance is reduced to 3.66 Å, versus 3.79 Å in the $I_h$ parent structure.

The $T_h$ symmetry group is the highest subgroup of the icosahedron. It is the epikernel of the distortion space of a fourfold degenerate $g_g$ coordinate.[6, 7] This $g_g$ mode is indeed one of the modes with negative curvature of the icosahedral instability. A similar distortion of $I_h$ to $T_h$ symmetry has been computed before for $Si_{60}$ and $Ge_{60}$ fullerenes.[8] However the characteristic feature of the $B_{80}$ case is the concomitant displacement of the capping atoms. The out-of-plane movement of the cap atoms is a very soft mode: as far as bond stretching interactions are concerned, the out-of-plane lifting of an atom which is in the centre of a coplanar net usually is described by a quartic potential, yielding a very shallow energy which explains the high mobility of the cap atoms. The valence π-electron structure of the boron buckyball is remarkably similar to buckminsterfullerene: the HOMO is a fivefold degenerate level of $h_u$ symmetry, the LUMO is the threefold degenerate $t_{1u}$ level, and the next LUMO is of $t_{1g}$ symmetry, exactly as in $C_{60}$. Suitable orbital components are presented in Fig. 2. Both HOMO and LUMO are almost fully localized on the $B_{60}$ orbit. The HOMO is π- bonding along the 6-6 bonds and antibonding along the 5-6 bonds. In the LUMO the bonding character is opposite.

In summary, we have found that the boron buckyball avoids the high symmetry icosahedral cage structure, and exhibits a spontaneous symmetry breaking to yield a puckered cage with only $T_h$ symmetry. It is one of the handfuls of molecules with this tetrahedral symmetry.[9, 10] The very shallow double well potential connecting the two alternate $T_h$ structures via an $I_h$ barrier will undoubtedly give rise to an interesting fluxional behaviour.

Furthermore it is found that the valence orbital structure of the boron buckyball is almost identical to the one in the carbon case. A far reaching implication would be that boron buckyball upon doping might exhibit metallic and even superconducting properties as well, which makes it an even more desirable target for chemical synthesis.

*Acknowledgements.* The authors are indebted to the Flemish Fund for Scientific Research (FWO-Vlaanderen) and the KU Leuven Research Council (GOA program) for continuing financial support.

# References

[1] N. G. Szwacki, A. Sadrzadeh, B. I. Yakobson, Phys. Rev. Lett. 98 (2007) 166804.

[2] Gaussian 03, Revision D.02, M. J. Frisch, et al., Gaussian, Inc., Wallingford CT, 2004.

[3] R. Ahlrichs, M. Bär, M. Häser, H. Horn and C. Kölmel, "Electronic structure calculations on workstation computers: the program system TURBOMOLE", Chem. Phys. Lett. 162 (1989) 165.

[4] L. J. Laaksonen, J. Mol. Graph. 10 (1992) 33.

[5] D. L. Bergman, L. Laaksonen, A. Laaksonen, J. Mol. Graph. Model. 15 (1997) 301.

[6] A. Ceulemans, L. G. Vanquickenborne, Structure & Bonding 71 (1989) 125.

[7] A. Ceulemans, P. W. Fowler, Phys. Rev. A 32 (1999) 155.

[8] B. X. Li, M. Jiang, P. L. Cao, Journal of Physics – Condensed Matter 11 (1999) 8517.

[9] J. C. G. Bunzli, S. Petoud, E. Moret, Spectroscopy Letters 32 (1999) 155.

[10] P.W. Fowler, P. Hansen, K.M. Rogers, S. Fajtlowicz, J. Chem. Soc. Perkin 2 7 (1998) 1531.

# Tables

**Table 1.** Calculated Total and Relative Energies (in parenthesis) of the three different $B_{80}$ structural isomers at HF/6-31G(d), B3LYP/6-31G(d) and B3LYP/SV(P) level.

| Symmetry/ State | Total Energies (hartree) and Relative Energies (kcal/mol, in parentheses) | | |
|---|---|---|---|
| | HF/6-31G(d)[a] | B3LYP/6-31G(d)[a] | B3LYP/SV(P) [a] |
| $I_h$ ($^1A_g$) | −1972.91937 (0) | −1987.55211 (0) | −1984.67339 (0) |
| $T_h$ ($^1A_g$) Isomer **A** | −1972.94659 (−17.08) | −1987.55290 (−0.50) | −1984.67623 (−1.78) |
| $T_h$ ($^1A_g$) Isomer **B** | −1972.93009 (−6.72) | −1987.55349 (−0.87) | −1984.67441 (−0.64) |

[a] Based on the optimized geometry at the indicated level. **A** and **B** have all real harmonic vibrational frequencies.

**Figure Captions**

**Figure 1.** Optimized geometries for $B_{80}$ buckyball at B3LYP/SV(P) and B3LYP/6-31G(d) levels. The $I_h$ symmetry structure is represented at the top; the distortions that accompany the $I_h$-$T_h$ structural change are represented using coloured arrows. Red arrows indicate displacement of the cap boron atoms towards the centre of the cage whereas green arrows indicate the same away from the centre of the cage, resulting in the $T_h$ symmetry isomer **A**. A reverse displacement will lead to the formation of isomer **B**. At the bottom the displaced atoms are represented in respective colours for both the isomers and bond lengths (in angstrom units) around the two different hexagons for Isomer **A**.

**Figure 2.** Shapes of the highest occupied and lowest unoccupied molecular orbitals (HOMO and LUMO) of the $I_h$ symmetric $B_{80}$ buckyball at B3LYP/SV(P) level in a pentagonal quantisation.

# Figures

**Figure 1.**

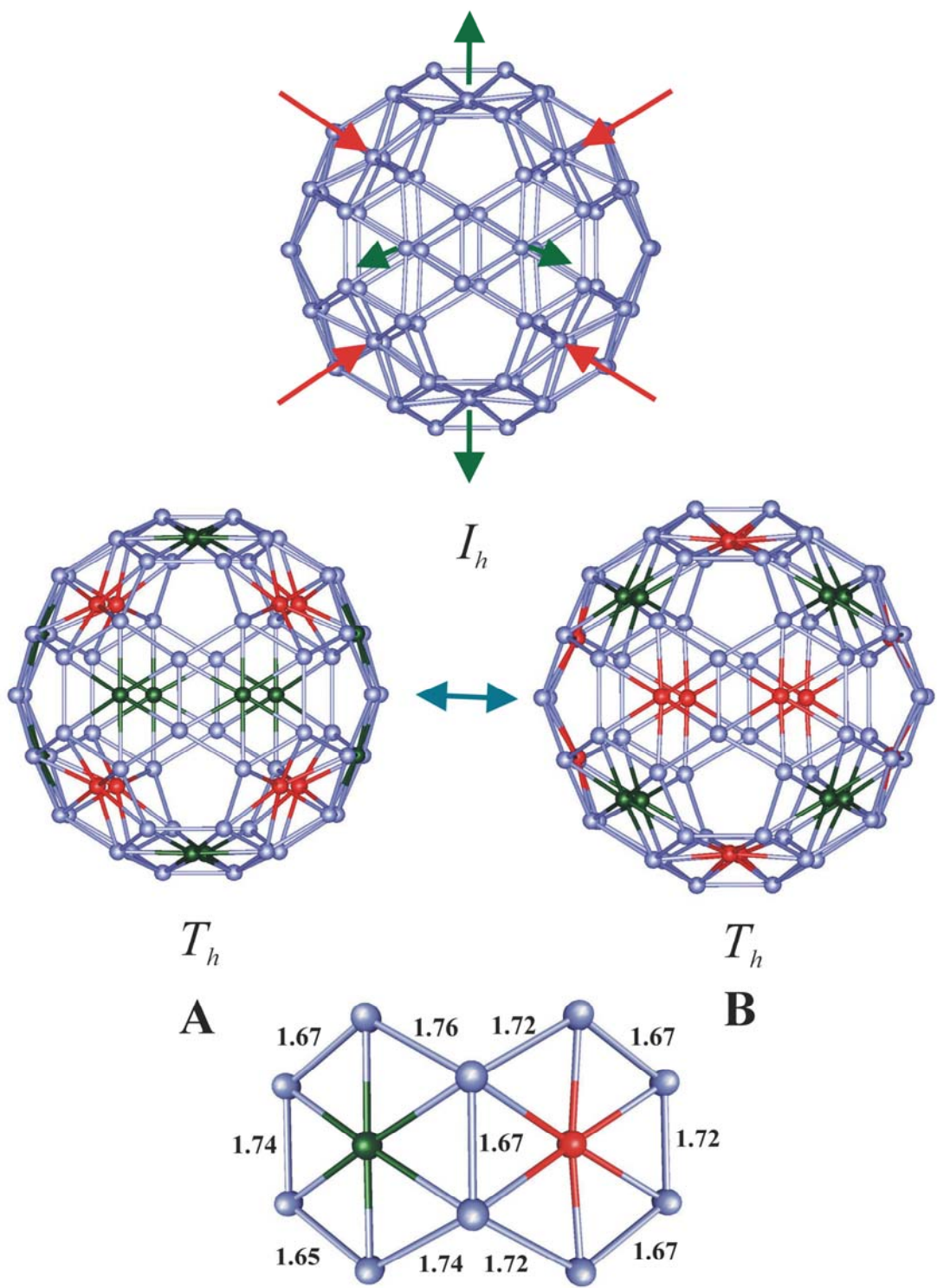

**Figure 2.**

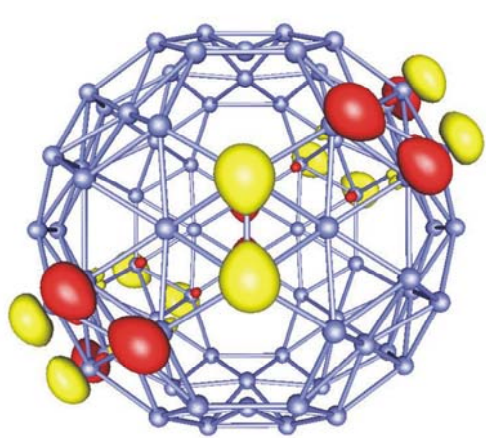
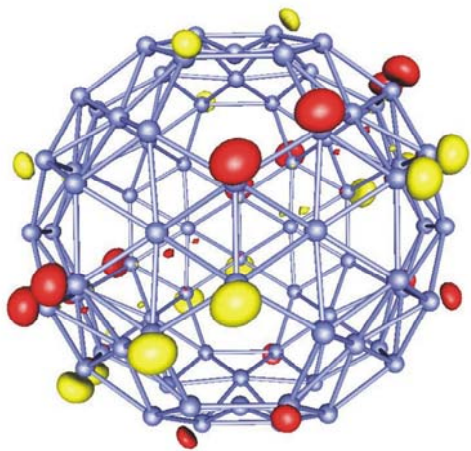

HOMO
6 H$_u$

LUMO
8 T$_{1u}$